\documentclass[a4paper,unsortedaddress,twocolumn,superscriptaddress,aip,jcp,reprint]{revtex4}

\usepackage[english]{babel}
\usepackage{natbib}

\usepackage{graphicx}

\usepackage{multirow}

\usepackage[T1]{fontenc}
\usepackage{lmodern}
\usepackage[colorlinks=true, pdfstartview=FitV, linkcolor=blue, 
            citecolor=blue, urlcolor=blue]{hyperref}
\usepackage{breakurl}
\usepackage{url}
\usepackage{bookmark}
\usepackage{epsfig}
\usepackage{microtype}
\usepackage[caption=false]{subfig}

\usepackage{epstopdf}
\usepackage{pstool}
\usepackage{braket}

\DeclareCaptionFormat{upper}{\MakeUppercase{#1}#2#3\par}
\usepackage[font=small, format=plain, textfont=small, up, justification=centerlast, position=below]{caption}
\usepackage{amsmath,amsfonts,amssymb}
\usepackage{float}

\newcommand{\dd}{{\rm d}}
\newcommand{\W}{\hat W}
\newcommand{\hS}{\hat S_\theta}
\newcommand{\Ht}{\hat H_\theta}
\newcommand{\ii}{\mathrm{i}}
\newcommand{\cprod}[2]{\left({#1}\right. |  \left. {#2} \right)}

\newcommand{\cket}[1]{\left| {#1} \right)}
\newcommand{\be}{\begin{equation}}
\newcommand{\ee}{\end{equation}}
\newcommand{\ba}{\begin{eqnarray}}
\newcommand{\ea}{\end{eqnarray}}

\newcommand{\lr}{\left(}
\newcommand{\rr}{\right)}
\newcommand{\vecr}{\vec{r}}

\begin{document}
\graphicspath{{figures/}}

\title{Accurate Complex Scaling\\ of Three Dimensional Numerical Potentials}

\author{Alessandro \surname{Cerioni}}
\email{alessandro.cerioni@esrf.fr}
\affiliation{European Synchrotron Radiation Facility, 6 rue Horowitz, BP220 38043 Grenoble Cedex 9, France}

\author{Luigi \surname{Genovese}}
\email{luigi.genovese@cea.fr}
\affiliation{Laboratoire de simulation atomistique (L\_Sim), SP2M, UMR-E CEA / UJF-Grenoble 1, INAC, Grenoble, F-38054, France}

\author{Ivan \surname{Duchemin}}
\affiliation{Laboratoire de simulation atomistique (L\_Sim), SP2M, UMR-E CEA / UJF-Grenoble 1, INAC, Grenoble, F-38054,
France}

\author{Thierry \surname{Deutsch}}
\affiliation{Laboratoire de simulation atomistique (L\_Sim), SP2M, UMR-E CEA / UJF-Grenoble 1, INAC, Grenoble, F-38054,
France}

\date{\today}

\begin{abstract}

The complex scaling method, which consists in continuing spatial coordinates into the complex
plane, is a well-established method that allows to compute resonant eigenfunctions of the time-independent Schr\"odinger operator.
Whenever it is desirable to apply the complex
scaling to  investigate resonances in physical systems defined on numerical
discrete grids, the most direct approach relies on the application of a similarity
transformation to the original, unscaled Hamiltonian.
We show that such an approach can be conveniently implemented in the Daubechies wavelet basis set, featuring a very
promising level of generality, high accuracy, and no need for artificial convergence parameters.
Complex scaling of three dimensional numerical potentials can be efficiently and accurately performed.
By carrying out an illustrative resonant state computation in the case of a one-dimensional model potential, we then
show that our wavelet-based approach may disclose new exciting opportunities in the field of computational non-Hermitian
quantum mechanics.

\end{abstract}

\maketitle

\section{Introduction}
\label{sec:intro}
Resonant states, sometimes referred to as metastable quasi-bound states,  are eigenstates of the
time-independent Schr\"odinger operator (atomic units are used throughout this work), 
\be
\hat H[\vecr] \equiv \hat T[\vecr] + \hat V[\vecr] = -\frac{1}{2}\nabla^2 +  V(\vecr)\;,\;\vecr\equiv(x,y,z)\,.
\ee
These solutions exhibit complex-valued eigenvalues, the (negative) imaginary part being inversely proportional to the resonance finite lifetime. Several equivalent definitions of resonant states can be formulated, each one suggesting a different
computational approach \cite{Hatano2010}. For instance, resonant states can be either regarded as poles of the
scattering matrix, or as solutions of the Schr\"odinger equation under the condition that only outgoing asymptotes exist
(Siegert boundary conditions \cite{PhysRev.56.750}).

When present, resonant states are not to be found within the Hermitian spectrum of $\hat H$, nor the
corresponding eigenfunctions are elements of the Hilbert space of square-integrable functions.
The fundamental works of Aguilar, Balslev, Combes and Simon \cite{Aguilar1971,Balslev1971,springerlink:10.1007/BF01649654},
proved that resonant states show up as discrete states in the spectrum of a non-Hermitian Hamiltonian operator obtained by scaling spatial coordinates by a complex factor, namely:
\be
\hat H[\vecr] \xrightarrow{\vecr\rightarrow \vecr\,e^{\ii\theta}} \hat H[\vecr e^{\ii\theta}]\equiv \Ht\,.
\ee
As soon as the complex scaling angle $\theta$ is greater than the critical angle $\theta_c \equiv
\arctan[|E_I/(E_R-E_{\textrm{th}})|]/2$, $E_{\textrm{th}}=V(\infty)$ being the threshold energy, the resonant state
corresponding to the eigenvalue $E = E_R - \ii E_I$, $E_I>0$, appears as a discrete eigenstate of $\Ht$, with
regular
eigenfunctions $\psi_\theta(\vecr)\rightarrow 0$ as $\vecr\rightarrow\infty$ (see Figs.~\ref{fig:DiaMolPreDis_Potential+Spectrum} and \ref{fig:psi_theta} for an illustration). 

A conspicuous number of articles contributed to
defining a non-Hermitian formalism in quantum mechanics. This formalism generalises all the fundamental ingredients of
traditional quantum mechanics (Rayleigh-Ritz variational principle, Hellmann-Feynman theorem, $\dots$), rephrased in
terms of a modified inner product (``$c$-product'' or bi-orthogonal product) defined as follows:
\be
\cprod{f}{g}\equiv \langle f^* | g \rangle \,.
\ee
Here, $\langle f | g \rangle$ represents the usual inner product in the $\mathbb L_2$ Hilbert
space.
For more comprehensive information, we refer to \emph{e.g.} Refs.~\onlinecite{moiseyev2011non,Moiseyev1998212,Klaiman2010,Junker1982207,doi:10.1146/annurev.pc.33.100182.001255,Ho1983,Moiseyev1984}.

Obtaining the complex-scaled version of the kinetic operator $\hat T$ is straightforward, since 
$\hat T_\theta = e^{-\ii 2\theta}\hat T$.
However, the evaluation of the complex rotated potential operator $\hat V_\theta$ is less obvious, especially if the potential is not known in closed form. For instance, this is always the case within self-consistent numerical simulations. To this end, different options have been proposed in the literature. Summarizing, they can be sketched out as follows: either the numerical potential is fitted with polynomials up to some order and then complex scaling is applied to the matrix elements in some finite basis set representation \cite{PhysRevA.20.814}, or the unscaled potential is represented on a complex-scaled basis set \cite{Ryaboy1995,museth:7008}. 

Complex-scaled Hamiltonians can also be obtained through a similarity transformation
\cite{Moiseyev1988}, without having to carry out the explicit evaluation of the original Hamiltonian in a complex
configuration space (see e.g. Ref.\ \onlinecite{moiseyev2011non}, p.\ 153). 
Within this approach, 
the complex-rotated Hamiltonian is obtained as 
\be
\label{eq:H_theta}
\Ht = \hS \hat H \hS^{-1}\,,
\ee
where the linear complex scaling operator $\hS$ is such that $\hat T_\theta = e^{-\ii 2\theta}\hat T$ and
\be\label{eq:SV}
[\hS \, V](\vecr) = V\lr \vecr\, e^{\mathrm{i}\theta}\rr\,.
\ee

The mapping between the $c$-orthonormal eigenfunctions of $\hat H$ and $\Ht$ reads $\cket{\psi} = e^{-\ii \theta/2}
\hS^{-1} \cket{\psi_\theta}$.
In the case of potentials for which a Taylor expansion is well-defined, the similarity transformation is \emph{strictly
equivalent} to the analytic continuation of the potential into the complex plane.

Different options for the operator $\hS$ are possible, as extensively discussed in Ref.\ \onlinecite{Moiseyev1988}. We
consider here the simplest formulation, namely:
\be\label{eq:S_operator}
\hS = e^{\theta \W}\,,\quad\mbox{with}\quad \hat{W} = \ii (\vec{r}-\vec{r}_0)\cdot\vec{\nabla}\,,
\ee
$\vec{r}_0=(x_0,y_0,z_0)$ being the fixed-point of the rotation. Clearly, $\hS^{-1} =\hat S_{-\theta}$ due to the group structure of the complex scaling transformation.

Although conceptually simple, implementing the complex scaling method through a similarity transformation in a
discrete basis set is not a trivial task.
A remarkable attempt can be found in
Ref.~\onlinecite{mandelshtam:6192}, where the similarity transformation operator
was represented in the sinc discrete variable representation (DVR) basis set. The accuracy of the method was ultimately proven to be somewhat sensitive to an artificial convergence parameter, depending on the
value of $\theta$ and on the basis set size. 

In this study, we show that Daubechies wavelets enjoy a number of features that allow obtaining a reliable representation of all the operators invoked by the complex scaling method,
with no need of artificial convergence parameters.

The plan of the paper is as follows: after briefly recalling the wavelet formalism, we show in
Secs.~\ref{sec:Hamiltonian_WL} and \ref{sec:S_WL}  how to represent the Schr\"odinger Hamiltonian and the
complex scaling operator in a wavelet basis set. In Sec.~\ref{sec:virial}, we review some of the standard results of
non-Hermitian quantum mechanics, which involve the complex virial theorem and eventually allow to assess the rate of
convergence towards exact bound and resonant states. In Sec.~\ref{sec:illustrative_examples}, we present the results
that we obtained when applying our wavelet-based algorithm to a number of model potentials, including a multi-centered
one-dimensional (1D) potential and a three-dimensional (3D) potential. We will show that the algorithm exhibits
excellent accuracy and numerical stability. As a further
confirmation, we will extensively discuss the computation of the lowest-lying resonances exhibited by a 1D model
potential, so as to prove that the complex virial theorem turns out to be fulfilled down to machine precision.
Our conclusions are summarized in Sec.\ \ref{sec:conclusion}, along with an outlook on future developments.

\section{Operators in Wavelet basis Sets}
\label{sec:Hamiltonian_WL}
The adoption of wavelets \cite{Goedecker2009} as a basis set in the present context is advisable for a number of reasons.  First, Daubechies wavelets~\cite{Daubechies1992Ten} present several properties which make them suitable for
the numerical simulation of isolated systems: they form a systematic, orthogonal and smooth basis, which is localized both in real and Fourier spaces and allows for adaptivity. Second, we will show that such functions will meet both the requirements of precision and localization found in many applications of the complex scaling method.
While referring the reader to Ref.~\cite{Goedecker2009} for an exhaustive presentation of how wavelet basis sets can be
used for numerical simulations, we here summarize the main properties of Daubechies wavelets, with a special
focus on the representation of the objects (wavefunctions and operators) involved in the present approach.

\begin{figure}
\includegraphics[width=0.48\textwidth]{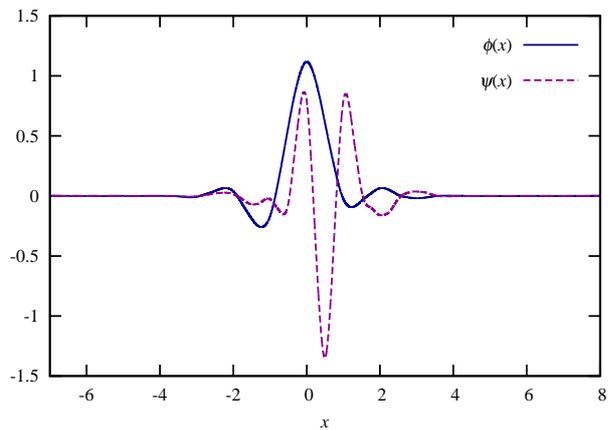}
\caption{Least asymmetric Daubechies wavelet family of order $2m=16$. Note that both the scaling function $\phi(x)$ and
the wavelet $\psi(x)$ are different from zero only within the interval $[1-m,m]$.\label{fig:Daub_16}}
\end{figure}

Every wavelet family comprises a \emph{scaling function} $\phi$, and a second function $\psi$ properly called \emph{wavelet}. Fig.~\ref{fig:Daub_16} illustrates the least asymmetric
Daubechies wavelet family of order $2m=16$, the basis set which is used in the present study.
These functions feature a compact support $[1-m,m]$ and are smooth, therefore localized in
Fourier space as well.

A basis set is simply generated by the integer translates of the scaling and wavelet functions, with arguments measured in units of the grid spacing $\ell$. For instance, a 1D domain of extension $L$, centered at $x=0$, can
be spanned by the following set of $N$ scaling functions,
\be\label{eq:WL_basis_set}
\braket{x|\phi_i}\equiv \phi_i(x) = \frac{1}{\sqrt{\ell}}\,\phi\left(\frac{x}{\ell}-i\right), \; i = -N/2,\dots, N/2\,,
\ee
where $\ell=L/(N-1)$ is the (uniform) grid spacing.
The basis set can be completed by the addition of the translates of the wavelet functions $\psi_i$.
These functions form a orthogonal basis set:
\be
\braket{\phi_i|\phi_j}= \delta_{ij}=\braket{\psi_i|\psi_j}\,,\quad\braket{\phi_i|\psi_j}= 0\,.
\ee

In three dimensions, a wavelet basis set can easily be obtained as the tensor product of one-dimensional basis
functions, combining wavelets and scaling functions along each coordinate of the Cartesian grid (see \emph{e.g.}
Ref.~\onlinecite{genovese:014109}).

The most important feature of any wavelet basis set is related to the concept of
\emph{multiresolution}. Such a feature builds upon the following \emph{scaling
equations} (or ``refinement relations''):
\ba
\label{eq:ref_rel}
\phi(x) &=& \sqrt{2} \sum_{j} h_j \,\phi(2x-j)\,;\\
\label{eq:ref_rel2}
\psi(x) &=& \sqrt{2} \sum_{j} g_j \,\phi(2x-j)\,,
\ea
which relate the wavelet representation at some resolution to that at twice the given resolution, and so on. According
to the standard nomenclature, the sets of the $h_j$ and $g_j=(-1)^j h_{1-j}$ coefficients are called \emph{low-} and
\emph{high-pass} filters, respectively. A wavelet family is therefore completely defined by its low-pass
filter. In the case of Daubechies-$2m$ wavelets, $j\in[1-m,m]$.

The representation $f(x)$ of a function in the above defined basis set is given by:
\be\label{eq:scalar_function_expansion}
f(x) = \sum_{i=-N/2}^{N/2} c_i\, \phi_i(x)+ \sum_{i=-N/2}^{N/2} d_i\, \psi_i(x)\,,
\ee
where the expansion coefficients are formally given by $c_i \equiv \braket{\phi_i|f}$, $d_i \equiv \braket{\psi_i|f}$. 
Using the refinement equations \eqref{eq:ref_rel} and \eqref{eq:ref_rel2}, one can map the basis appearing in Eq.~\eqref{eq:scalar_function_expansion} to an equivalent one including only scaling functions on
a finer grid of spacing $\ell/2$.

The multiresolution property plays a fundamental role also for the wavelet representation of differential operators. For example, it can be shown   that the
\emph{exact} matrix elements of the kinetic operator,
\be
T_{i-j}\equiv-\frac{1}{2}\int \dd x \, \phi_i(x) \partial^2 \phi_j(x)\,,
\ee
equal the entries of an eigenvector of a matrix which solely depends on the  low-pass filter (see \emph{e.g.} Ref.~\onlinecite{Goedecker2009}). The
construction of the kinetic operator $\hat T$ in a wavelet basis set does not require any sort of approximation or numerical
integration.  In the next section, we will show that the matrix elements of $\hat W$ can be calculated
in a similar way.

The discretization error due to Daubechies-$2m$ wavelets is controlled by the grid spacing. Daubechies-$2m$ wavelets exhibit
$m$ vanishing moments, thus any polynomial of degree less than $m$ can be represented exactly by an expansion over the
sole scaling functions of order $m$. For higher order polynomials the error
is $\mathcal{O}(\ell^m)$, \emph{i.e.}\ vanishingly small as soon as the grid is sufficiently fine. 
Hence, the difference between the representation of Eq.~\eqref{eq:scalar_function_expansion} and the exact function
$f$ decreases as $\ell^m$.
Among all the wavelet families, Daubechies wavelets feature the minimum support length
for a given number of vanishing moments.

Given a potential $V$ known numerically on the points $\{x_k\}$ of a uniform grid, 
it is possible to identify an effective approximation for the potential matrix elements $V_{ij}\equiv \bra{\phi_j} V
\ket{\phi_i}$.
It has been shown \cite{Neelov2006312,genovese:014109} that a quadrature filter $\{\omega_{k}\}$ can be defined such that the matrix elements given by
\be\label{eq:magic_filter_application}
V_{ij}\equiv \bra{\phi_j} V \ket{\phi_i} = \sum_{k}  \omega_{k-i}\, V(x_k)\,\omega_{k-j}\,,
\ee
yield excellent accuracy with the optimal convergence rate $\mathcal O(\ell^{2m})$ for the potential energy.
The same quadrature filter can be used to express the grid point values of a (wave)function given its
expansion coefficients in terms of scaling functions:
\begin{align}
f(x_k) &= \sum_{i} c_i\, \omega_{k-i} + \mathcal O(\ell^{m})\,;\\
c_i &= \sum_{k} f(x_k)\, \omega_{k-i} + \mathcal
O(\ell^{m})\,.
\end{align}
As a result, the potential energy can equivalently be computed either in real space or in the wavelet space, \emph{i.e.}\ $\bra f V \ket f= \sum_k f(x_k) V(x_k) f(x_k) \equiv \sum_{ij}c_i V_{ij} c_j$.
The quadrature filter elements can therefore be considered as the most reliable transformation between grid point values
$f(x_k)$ and scaling function coefficients $c_i$, as they provide exact results for polynomials of order up to $m-1$ and
 do not alter the convergence properties of the basis set discretization.
The filter $\{\omega_k\}$ is of length $2m$ and is defined unambiguously by the
moments of the scaling functions (which in turn depend only on the low-pass filter) \cite{230060,johnson:8309}.

Using the above formulae, the Hamiltonian matrix $H_{ij}=T_{ij}+V_{ij}$ can be constructed. Note that, in contrast to other discretization schemes (finite differences,
DVR, plane waves, \emph{etc.}), in the wavelet basis set \emph{neither} the potential \emph{nor} the kinetic terms have
diagonal representations. Instead, $\hat H$  is represented by a \emph{band matrix} of width $4m-1$.

\begin{figure}
\centering
\includegraphics[width=0.48\textwidth]{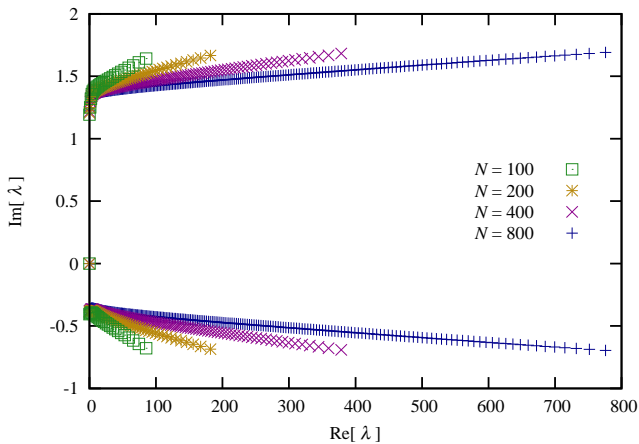}
\medskip
\caption{Spectrum of the operator $\hat W$ - cf.\ Eq. \eqref{eq:W_operator} - for different sizes $N$ of the basis set. Knowing that the spectrum of $\W$ is symmetric around the imaginary axis, only the right half-plane is shown.
\label{fig:spectrum_of_W}}
\end{figure}

\section{Complex Scaling via Similarity Transformation}
\label{sec:S_WL}

The complex scaling operator $\hS$ being separable along the three spatial directions, and the wavelet transform of 3D objects being given by the tensor product of three 1D wavelet transforms, it is sufficient to build the complex scaling operator suitable to working on 1D input functions. 3D complex-scaled potentials can thus be obtained through the threefold application of the same 1D operator, each time acting along a different direction:
\begin{equation}
 V(x\,e^{\ii\theta}, y\, e^{\ii \theta}, z\, e^{\ii \theta}) = [{\hS}^x {\hS}^y {\hS}^z V](x,y,z)\,.
\end{equation}
In what follows, we drop the superscript denoting the spatial direction and focus on the construction  of the 1D complex scaling operator, taking 
\be\label{eq:W_operator}
\W = \ii(x-x_0)\frac{\partial}{\partial x}\,.
\ee

Owing to the scaling properties of wavelets and scaling functions, the
matrix elements $W_{i,j}$ of the operator $\W$ can be computed exactly. The final result reads as 
\be
W_{i,j} = \ii\left[w_{i-j}+\left(\frac{i+j}{2}-x_0 \right)d_{i-j}\right]\,.
\ee
The definitions of the filter coefficients $\{w_j\}$ and $\{d_j\}$ are deferred to
Appendix~\ref{app:matrix_elements_of_W}, along with the details of their computation. Let us note that, within the Daubechies-$2m$ family, $W_{i,j}=0$ for any pair
$\{i, j\}$ such that $|i-j|>2m-2$. Moreover, being $\W$ scale-invariant, its spectrum depends only on the size $N$ of
the basis set.
Fig.~\ref{fig:spectrum_of_W} shows that the eigenvalues of $\hat W$ span a range which reaches a maximum that
is $\lesssim N$.
 
Once the \emph{exact} wavelet representation of the operator $\hat W$ is available, the operator $\hS$ can be 
represented by its spectral decomposition over the eigenvalues and eigenvectors of 
$\W$: 
\be\label{eq:spectral_decomposition}
\hS = e^{\theta \hat W} = \sum_{i=1}^{N} e^{\theta\lambda_i}\,  | \psi_{\lambda_i}\rangle \langle
\psi_{\lambda_i} | \,,
\ee
where $\lambda_i$ is the eigenvalue corresponding to the eigenvector $|\psi_{\lambda_i}\rangle$. However, special
attention has to be paid when using Eq.\ \eqref{eq:spectral_decomposition} as is, since
numerical
instabilities are likely to occur depending on the maximum value attained by $\textrm{Re}(\theta\lambda_i)$
(in other words: on the condition number of the operator $\hS$). For instance, if the computation is carried out in
double-precision floating point arithmetic (by far the most typical case), the relevant terms of the spectral
 decompositions \eqref{eq:spectral_decomposition} should lie in the region $10^{-15} <
|e^{\theta\lambda_i}| < 10^{15}$. 
In order for the complex rotation to be
reliable, the scalar product $\langle \psi_{\lambda_i} | V \rangle$ should be non-zero only for a subset $\{\lambda_i\}$ for which  $|e^{\theta\lambda_i}|$  falls within the same range. 
This is however likely to happen, provided the potential $V$ is sufficiently localized in real and reciprocal
space. Actually, since the exact eigenfunction of $\hat W$ corresponding to the (complex)
eigenvalue $\lambda$ is given by $\psi_{\lambda}(x) \propto (x-x_0)^{-\ii \lambda}$, the real part of
$\lambda$ can be deemed as a wavenumber in a logarithmic configuration space. 
Given that $x\, \partial_x =-(1 + p \, \partial_p)$, the same behavior would hold also in reciprocal
space, $\psi_{\lambda}(p) \propto p^{\ii \lambda-1}$.
Smooth and localized functions are
therefore likely to exhibit non-trivial projections only onto the lowest-lying eigenstates of $\hat W$.

At the same time, for two functions $f(x)$ and $g(x)=f(x)+c$ that differ by a global constant $c \in \mathbb C$, the
equality $\hS g = \hS f + c$ is expected to hold for any $\theta$. The numerical representation of the operator
$\hS$ should namely behave as the identity on a generic constant vector $|c\rangle$.
A numerically stable implementation should therefore be such that $\langle \psi_{\lambda_i} | V \rangle= \langle
\psi_{\lambda_i} | V +c\rangle$ for any $\lambda_i \neq 0$.
 Such a requirement can be enforced by imposing suitable boundary conditions on the matrix elements of $\W$, as
demonstrated by the following result, which involves the $n$-th moment of a generic left-eigenfunction $\psi_\lambda(x)$
of $\W$ (assuming $x_0=0$, $\lambda \psi_\lambda(x) \equiv \psi_\lambda(x) \overleftarrow{W} = \ii\, \partial_ x \left[x\, \psi_\lambda(x) \right]$): 
\begin{multline}
(\lambda +\ii\, n)\int_{-L/2}^{L/2}  \dd x\,\psi_\lambda (x)  x^{n}  = \\
\ii\left(\frac{L}{2}\right)^{n+1} \left[\psi_\lambda(L/2) + (-1)^{n}\psi_\lambda(-L/2) \right] \;.
\end{multline}
In particular, for $n=0$ the boundary term on the right hand side becomes zero if $\psi_\lambda(L/2)=
-\psi_\lambda(-L/2)$, and $ \langle\psi_\lambda | c \rangle~=~0$ for any $\lambda \neq 0$. In other words, the only left-eigenvector of $\W$ that has a non-trivial projection on a
constant, is the one corresponding to the zero eigenvalue. 
Such boundary conditions can be imposed by setting 
\be
W_{i,j\pm N} = -\ii\left[w_{i-j}+\left(\frac{i+j}{2}-x_0 \right)d_{i-j}\right]\,.
\ee
The zero eigenvalue can actually be
found in the spectrum of the $\W$ operator that is so obtained (see Fig.\ \ref{fig:spectrum_of_W}), the pair matching
right-eigenvector being identically constant.

\begin{figure*}
\includegraphics[width=\textwidth]{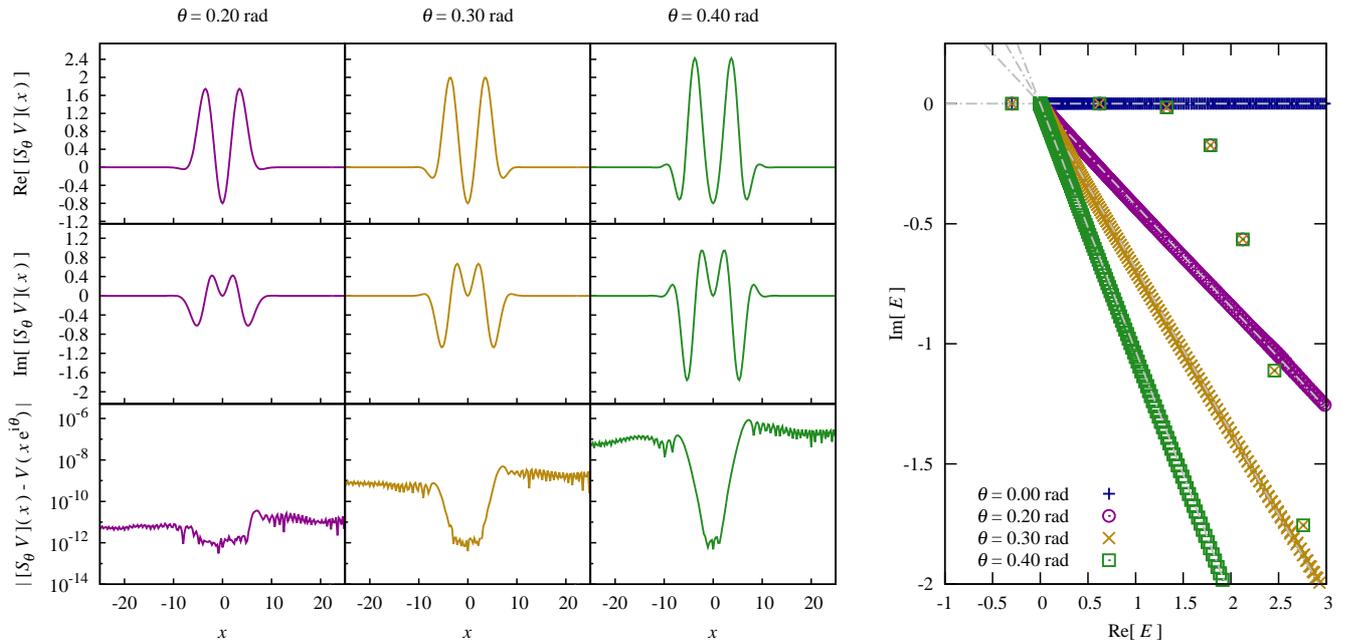}
\caption{
\label{fig:DiaMolPreDis_Potential+Spectrum}
Left: the results of the application of the wavelet-based complex scaling method to the potential \eqref{eq:pot} for different complex scaling angles. In the bottom row, we plot the absolute value of the difference between the exact complex-scaled potential $V(x e^{\ii \theta})$ and that obtained through similarity transformation in the wavelet basis set, $ [\hS V](x)$.
Right: the low-energy portion of the spectrum of the corresponding complex-scaled Hamiltonian. One can note that continuum states align along rays (dot-dashed straight lines in the plot) that are rotated by $2\theta$ with
respect to the real axis, while bound and resonant states can be identified by the fact that they do not depend on $\theta$.}
\end{figure*}

\section{The Complex Virial Theorem}
\label{sec:virial}
As a result of the Balslev-Combes theorem \cite{Balslev1971}, bound and resonant eigenstates are expected to be stationary with respect to
variations of the complex scaling angle $\theta$.
Moreover, the complex analog of the Hellmann-Feynmann  theorem \cite{moiseyev2011non} allows to relate the variation of
the $i$-th eigenvalue $E^{(i)}$ with respect to $\theta$ (considered as a variational parameter) to the quantum expectation
value of $\partial \Ht / \partial\theta$ on the corresponding eigenstate $|{\psi_\theta^{(i)}})$:
\be
\frac{\dd E^{(i)}}{\dd \theta} = \left(\psi_\theta^{(i)} \right| \frac{\partial \Ht}{\partial \theta}\left
|\psi_\theta^{(i)} \right),
\ee
being 
\be
E^{(i)} = \frac{(\psi_\theta^{(i)} | \Ht | \psi_\theta^{(i)})}{(\psi_\theta^{(i)}|\psi_\theta^{(i)})}\; .
\ee
Given the definition of $\Ht$ in Eq.\ \eqref{eq:H_theta}, it is easy to show that
\be
\frac{\partial \Ht}{\partial \theta} = [\hat W, \Ht]\,,
\ee
and to rephrase the requirement of $\theta$-independence in terms of the complex virial theorem (CVT):
\be
\frac{\dd E^{(i)}}{\dd \theta} = 0 \Longleftrightarrow (\psi_\theta^{(i)} | [\hat W, \Ht] | \psi_\theta^{(i)}) = 0\,.
\ee 
As shown \emph{e.g.} in Ref.\ \onlinecite{doi:10.1021/jp9607881} it is possible to prove that
\be\label{eq:virial2}
[\hat W, \Ht] = -2\ii\, (\hat T_\theta +  \hat U_\theta)\,,
\ee
where
\be\label{eq:U}
\hat U \equiv -\frac{x-x_0}{2}\frac{\partial \hat V}{\partial x} = \frac{\ii}{2} \W \hat V\,.
\ee
Eventually, the virial theorem reads as follows:
\be
\label{eq:virial_test}
\left(\psi^{(i)}_\theta \right| \hat T_\theta + \hat U_\theta \left|  \psi^{(i)}_\theta \right) = 0\,.
\ee
When using finite basis sets, the condition \eqref{eq:virial_test} may not be satisfied, or be satisfied to a great extent only within
certain intervals of the angle $\theta$. Such an occurrence is extensively reported in the literature \cite{YarisWinkler1}. Resonant eigenvalues have been shown to move along the so-called $\theta$-trajectories in the complex plane as $\theta$ is varied, the features of which (cusps, in particular) can be used to assess the level of convergence towards stationary
points, and might hint at the selection of the optimal complex scaling angle
\cite{moiseyev:4739,doi:10.1080/00268977800102631,PhysRevA.20.814}. 

The evaluation of $\dd E^{(i)}/\dd \theta$ thus represents an assessment of the degree of convergence towards true bound and resonant states in numerical computations with finite basis sets.
A closer look at Eq.\ \eqref{eq:U}  reveals that no additional effort is required to obtain the operator $\hat U$ in the
wavelet basis set, once $\W$ has been generated. Moreover, the complex-scaled counterpart $\hat U_\theta$
can be obtained by acting on $\hat U$ with the complex scaling operator $\hS$, exactly as if $\hat U$ were a generic
potential operator.

\begin{figure*}
\includegraphics[width=\textwidth]{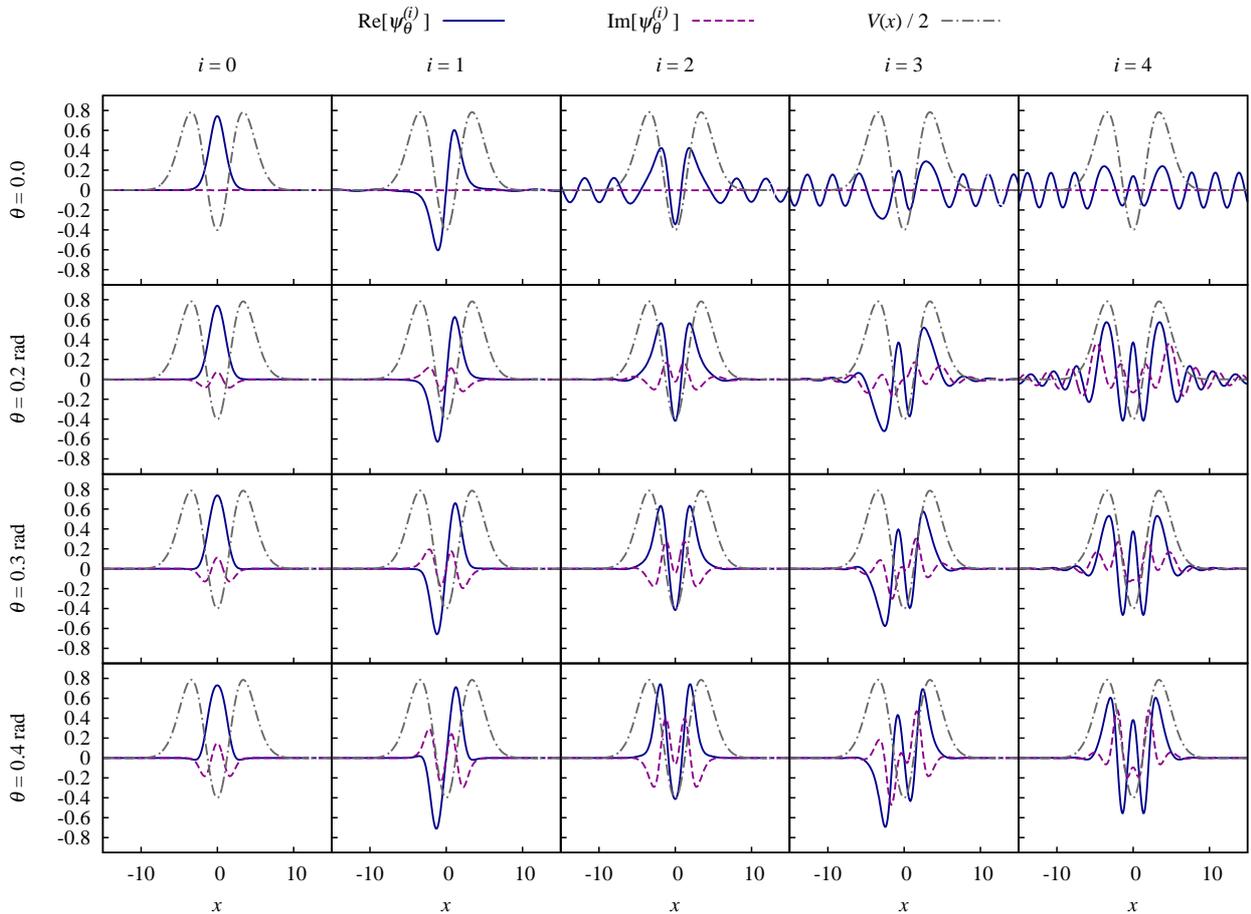}
\caption{\label{fig:psi_theta} Complex-rotated $c$-normalized eigenfunctions corresponding to the five lowest-lying eigenstates (from
left to right: one bound state and four resonant states), computed at $\theta = \{0, 0.2, 0.3, 0.4\}\;\mbox{rad}$. Solid (dashed) lines correspond to the real (imaginary) part of the eigenfunctions. Dot-dashed
lines represent the potential (rescaled by a numerical factor), which is superimposed so to
highlight the localization of the complex-rotated eigenfunctions inside the interaction region.}
\end{figure*}
\begin{figure}
\includegraphics[width=0.48\textwidth]{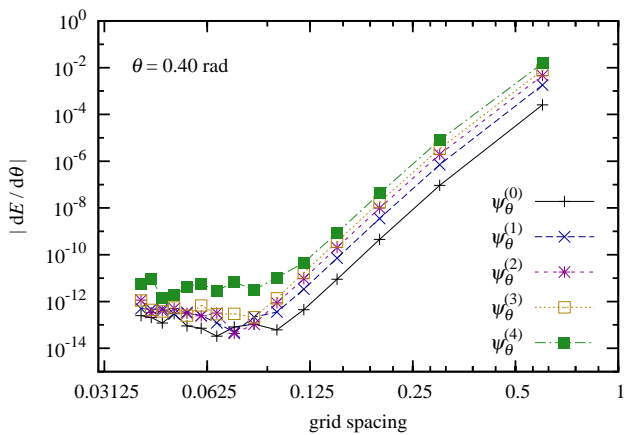}
\caption{\label{fig:CVT_vs_h} Measurement of the fulfillment degree of the complex virial theorem, Eq.~\eqref{eq:virial_test}. 
The result of the
evaluation of Eq.\ \eqref{eq:virial_test} is shown as a function of the grid spacing of the simulation domain,  for the five lowest-lying states of the complex-scaled Hamiltonian at $\theta
= 0.4\;\textrm{rad}$ obtained through similarity transformation in the Daubechies-16 wavelet basis set. The potential is that of Eq.~\eqref{eq:pot}. }
\end{figure}
\begin{figure}
\centering
\includegraphics[width=0.48\textwidth]{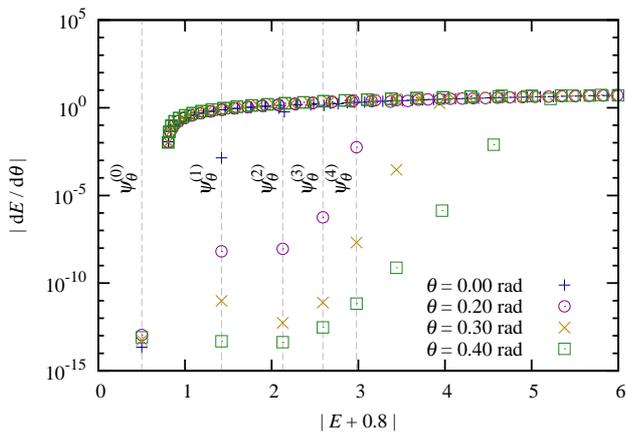}
\bigskip
\caption{The results obtained for $\Ht$ with the potential as from Eq.~\eqref{eq:pot}, with a focus on the low-energy end. The grid spacing was set equal to $0.075$ (cf.\ Fig.~\ref{fig:CVT_vs_h}). The offset $0.8$ ($=-V(x=0)$) was added to the eigenenergies in order to avoid ambiguities in the ordering of states, even in terms of the absolute values of the associated eigenenergies. Vertical dashed lines are superimposed on the eigenenergies of the five lowest-lying discrete states.
\label{fig:virial_vs_E}}
\end{figure}

\section{Illustrative examples}\label{sec:illustrative_examples}

In order to validate our wavelet approach to complex scaling via similarity transformation, we first discuss the application of our method to a specific 1D model potential. We chose to work on the following potential,
\be\label{eq:pot}
V(x) = \left(\frac{1}{2}x^2 - a\right) e^{-\lambda x^2} + E_{\textrm{th}}\,,
\ee
with $a = 0.8$, $\lambda = 0.1$ and $E_{\textrm{th}}=0$, because the same potential, with $E_{\textrm{th}} = 0.8$,  has been thoroughly studied in the literature, hence we could cross-check our results against the already published ones. Let us point out that we do not aim to provide new information concerning the resonances exhibited by the model potential \eqref{eq:pot}, instead to show that 1.\ the wavelet-based complex scaling transformation is extremely accurate; 2.\ the finite wavelet basis set computation of resonant states yields very good figures of the fulfillment of the CVT.

The potential \eqref{eq:pot}, which models diatomic molecular
pre-dissociation resonances, has
already been studied in a number of publications, namely in 
Refs.\ \onlinecite{doi:10.1080/00268977800102631,moiseyev:3623} (adopting a Gaussian basis set), 
in Refs.\ \onlinecite{PhysRevA.24.1636,PhysRevA.26.1804} (employing Weyl's theory and the direct numerical integration
of the complex-rotated Schr\"odinger equation under Siegert boundary conditions), in Refs.\
\onlinecite{0022-3700-15-1-008,PhysRevA.26.1802} (employing a non-Hermitian generalization of the Milne's method), in Ref.\ \onlinecite{0953-4075-31-7-009} (using a complex absorbing potential and plane waves as basis
functions) and in Refs.\ \onlinecite{museth:7008,bludsky:1201,leforestier:1203,Fang2008,Fang2008a}
(using the sine DVR). More specifically, in Refs.\  \onlinecite{Fang2008,Fang2008a} the
potential is complex-scaled while the basis functions are not; in Refs.\  \onlinecite{museth:7008},
\onlinecite{bludsky:1201} and \onlinecite{leforestier:1203} the basis functions are scaled, but not the potential. The model potential \eqref{eq:pot} exhibits only one bound state (at $E-E_{\textrm{th}}\simeq -0.3$), and several
resonant states. 

The outcome of the complex scaling operation applied to the potential \eqref{eq:pot} is shown in Fig.\
\ref{fig:DiaMolPreDis_Potential+Spectrum}, together with a comparison against the exact complex-scaled potential. An
excellent agreement is found throughout the spatial range for all the values of $\theta$. 

The full diagonalization of the complex-scaled Hamiltonian obtained through similarity transformation provides a further confirmation of the reliability of the wavelet-based complex scaling operation. To this effect we show, in the right panel of Fig.~\ref{fig:DiaMolPreDis_Potential+Spectrum}, the low-energy region of the $\Ht$ spectrum and, in Fig.\ \ref{fig:psi_theta}, the eigenfunctions of the five lowest-lying states (one bound state and four resonances).
It can be observed that resonant eigenfunctions are well localized inside the interaction region and
well-behaved throughout the spatial range, pretty much like a bound state. The localization increases while increasing
the complex scaling angle (the effect being more evident for the resonances at higher energy). Also shown are the
eigenfunctions sorted out from the spectrum of $\hat H_{\theta=0}$ by
looking up the eigenvalues which were the closest to the resonant eigenvalues computed at $\theta>0$.

In Fig.~\ref{fig:CVT_vs_h}, the rates of convergence towards true resonant states (namely, the degree of fulfillment of the CVT) are displayed. The rates improve upon decreasing the grid spacing, as a consequence of the systematicity of Daubechies wavelets. In the case of a sufficiently fine grid, the figures of the violation of the CVT become as little as $\lesssim 10^{-11}$, for all five lowest-lying states. Such figures have to be compared with those referring to continuum states ($\gtrsim 10^{-2} $, see Fig.\ \ref{fig:virial_vs_E}). The gap of several orders of magnitude between bound/resonant states and continuum states allows to easily distinguish the former from the latter within the $\Ht$ spectrum. It is also interesting to note how the measure of the violation of the CVT changes \emph{vs}\ $\theta$ for the different resonances.

\begin{figure}
\includegraphics[width=0.48\textwidth]{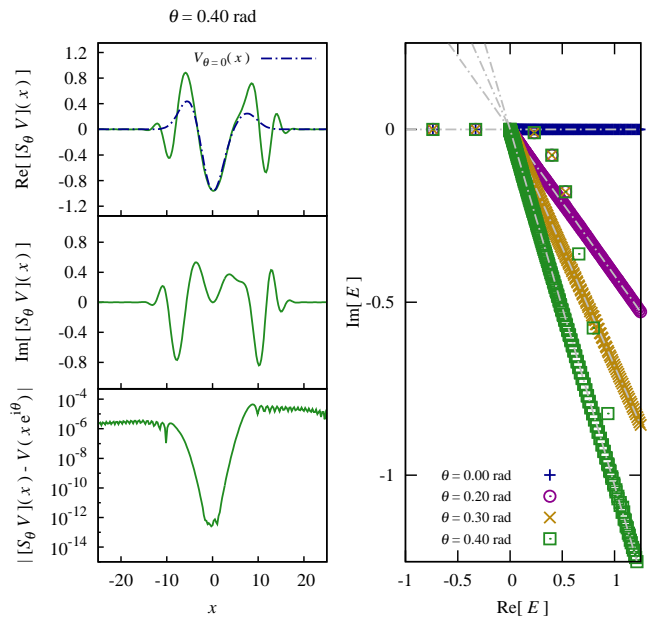}
\caption{\label{fig:AsymPot} The result of the wavelet-based complex scaling as applied through similarity transformation to an illustrative multi-centered potential $V(x) = 0.25 \exp [-0.1(x-7.5)^2] + 0.5 \exp [-0.1(x+5)^2] - \exp(-0.1 x^2)$. Left: real (top) and imaginary (center) part of the complex-scaled potential at $\theta=0.4\;\textrm{rad}$, and comparison (bottom) against the exact result. Right: low-energy portion of the spectrum of the corresponding non-Hermitian Hamiltonian. By virtue of the $\theta$-independence one can identify a pair of bound states and a number of resonances.}
\end{figure}

\begin{figure}
\includegraphics[width=0.49\textwidth]{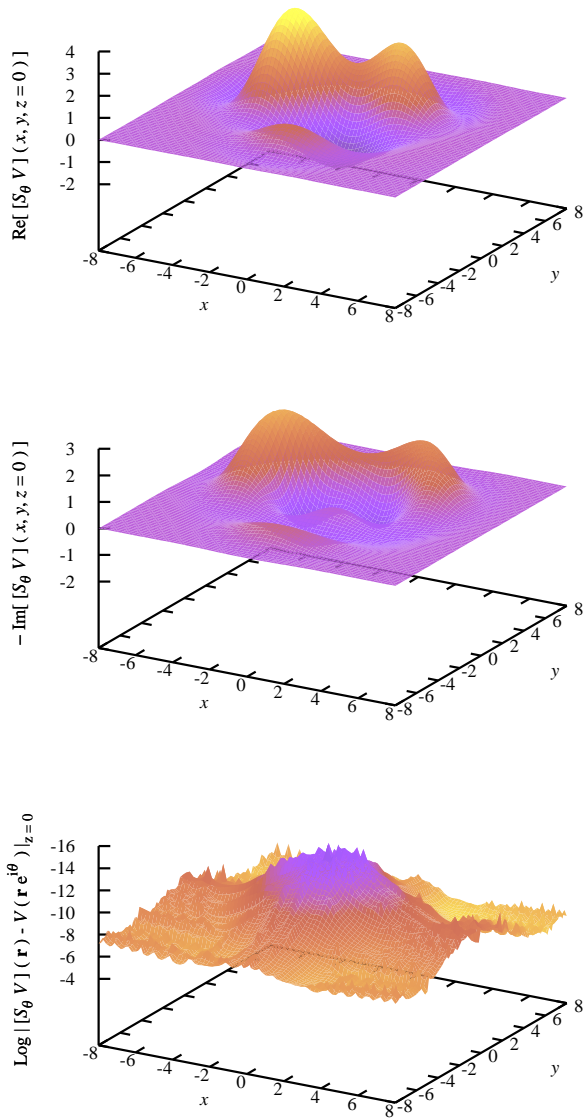}
\caption{\label{fig:3D} The result of the wavelet-based complex scaling applied to a fictitious 3D potential, $V(\vec{r}) = \sum_{k = 0}^3 a_k e^{-\beta_k |\vec{r}-\vec{r}_k|^2}$ with $a_0 = -3$, $a_1 = 2$, $a_2 = 4$, $a_3  = 3$, $\beta_0 = 0.1$, $\beta_1=\beta_2=\beta_3 = 0.2$, $\vec{r}_0 = 0$, $\vec{r}_1 = 3(0, -1, 1/2)$, $\vec{r}_2 = 3/2(-\sqrt{3}, 1, 0)$, $\vec{r}_3 = 3/2(1, \sqrt{3},0)$ and $\theta = 0.3\;\textrm{rad}$: 2D surfaces at $z=0$
representing the real (top), imaginary (center) part of $[\hat S_\theta V](r)$  and the discrepancy (bottom) from the exact result. Note that, for better visibility, the opposite of the imaginary part is plotted. }
\end{figure}

We wish now to discuss the results obtained upon applying the method to a multi-centered one-dimensional potential (Fig.\ \ref{fig:AsymPot}) and to a 3D potential (Fig.\ \ref{fig:3D}), with the twofold aim of providing evidence that the method also works in three dimension and that no special restrictions have to be imposed on the symmetry of the potential. While referring to the figure caption for further information on the shape of the two potentials, we wish to remark that the wavelet-based complex scaling method proves to be, once again, very accurate.  

We look forward to applying the wavelet-based complex scaling method to genuine \emph{ab initio} potentials, as a first crucial step towards resonances in realistic 3D (isolated) systems. 

\section{Conclusion}
\label{sec:conclusion}

We have shown that the adoption of wavelets as a basis set allows to implement the complex scaling method via similarity
transformation in a rigorous and effective way.
Since wavelets display well-defined transformation properties upon rescaling of the spatial coordinates, they are
especially well-suited for the discretization of scale-invariant operators such as the complex scaling generator $\hat
W$.
In addition, the localization of wavelets (both in real and
reciprocal space) allows to represent complex scaling transformation through the spectral decomposition over the
eigenstates of a matrix that is \emph{inherently} band-like. 
As a consequence, no artificial convergence parameter has to be
introduced. 

By testing the method on a host of 1D and 3D model potentials, we were
able to prove that our implementation is also very accurate. 
This general technique, starting from the numerical values
$\{f(x_k)\}$ of a generic function $f(x)$
on a (uniform) real space grid,  computes the values $\{f(x_k e^{\ii \theta})\}$ by performing a similarity
transformation in the Daubechies
wavelets basis set. The output may be further processed in any other basis set of choice, \emph{e.g.}\ for
the extraction of resonant states of the complex-scaled Hamiltonian.

By carrying out a resonant state computation in the case of a
one-dimensional model system, we were able to demonstrate that our approach exhibit excellent convergence rate in terms
of the size of the basis set. The violation of the
complex virial theorem can become comparable to the machine precision, hence totally negligible.

Resonant states are supposed to provide a concise description of one-particle excitations, meaning that a
certain energy range ought to be spanned by \emph{a few resonant states} rather than \emph{a virtually infinite set of continuum
states}.
It has been claimed, for instance, that ``\emph{one can construct the Green function simply in the form of a sum over
the Siegert states, avoiding the annoying integral over the continuum}.'' (citation from Ref.\
\onlinecite{PhysRevLett.79.2026}; see also Refs.~\onlinecite{Hatano2010,Hatano2010a}). Several other
investigations indicate the possibility of describing scattering cross sections, optical absorption spectra \emph{et
cetera} by means of a collection of purely discrete (bound and resonant) states \cite{PTP.99.801}.
Nevertheless, the very numerical computation of resonant states is a formidable challenge. In our study, we showed that wavelets allow obtaining a complex-scaled Hamiltonian which can
contain resonant states in its spectrum (if any). Irrespectively of the basis set in which one wishes to further the
computation, in realistic numerical simulations the size of the Hamiltonian matrix can be so large that the full diagonalization becomes practically unaffordable. Moreover, one has to cope with the fact that, within the
framework of non-Hermitian quantum mechanics, the variational principle states that
``true'' states are only \emph{stationary} points along the variational trajectories (instead of \emph{minima}). Hence,
methods that are already in use for the computation of bound states cannot be deployed as such. 

A crucial achievement
toward the numerical computation of resonant states should regard the development of clever techniques for their
direct extraction, allowing to focus on selected sub-regions of the spectrum of the
non-Hermitian Hamiltonian (like, for instance, the filter diagonalization method \cite{Santra20021}). Work is in progress
in this direction. We also look forward to carrying out computations in the frame of many-body perturbation theory \emph{with} the inclusion of resonant states \cite{ANDP:ANDP201200062}, hoping to be able to shed new light on the
theoretical description of optical and electronic excitations. To this respect, the present study represents a first essential step, since it opens up the possibility of obtaining the complex-scaled version of Hamiltonians based on 3D  \emph{ab initio} potentials.


\acknowledgments
The authors wish to thank Eric Canc\`es and Salma \mbox{Lahbabi} for valuable discussions and 
Claudio Ferrero for the critical proofreading. A.C. acknowledges the financial support of the French National Research Agency in the frame of the
``\textsc{NEWCASTLE}'' project.

\appendix
\section{Representation of the $\hat W$ operator in a wavelet basis set}
\label{app:matrix_elements_of_W}
Computing the wavelet representation of the operator $\W$ - cf.\ Eq.\ \eqref{eq:W_operator} - amounts to evaluating the following matrix elements:
\begin{subequations}
\label{eq:W_matrixElements}
\ba
W^{(00)}_{m,n} &\equiv& \int_\mathbb{R} \dd x\, \phi_m(x) \hat W \phi_n(x)\;;\\
W^{(01)}_{m,n} &\equiv& \int_\mathbb{R} \dd x\, \phi_m(x) \hat W \psi_n(x)\;;\\
W^{(10)}_{m,n} &\equiv& \int_\mathbb{R} \dd x\, \psi_m(x) \hat W \phi_n(x)\;;\\
W^{(11)}_{m,n} &\equiv& \int_\mathbb{R} \dd x\, \psi_m(x) \hat W \psi_n(x)\;,
\ea
\end{subequations}
where $\phi_i(x)\equiv \phi(x-i)$ and $\psi_i(x)\equiv \psi(x-i)$ are the scaling function and the wavelet belonging to a wavelet family, respectively. In our specific implementation, we adopted Daubechies wavelets, for the reasons outlined in Sec.\ \ref{sec:Hamiltonian_WL}.
Introducing the following variables,
\be
I \equiv \frac{m+n}{2}\,,\qquad J \equiv \frac{m-n}{2}\,,
\ee
and switching from $x$ to $x'= x-I$, we can split $W^{(00)}_{m,n}$ into a translation-invariant part, which indeed
depends only on $J$, and a non-translation-invariant part. Eventually,
\be
W^{(00)}_{m,n}  = \ii\left[w^{(00)}_{m-n}+\left(\frac{m+n}{2}-x_0\right)\, d^{(00)}_{m-n} \right]\,,
\ee

where
\be\label{eq:a_J_def}
w^{(00)}_{i} \equiv \int_\mathbb{R} \dd x \, \phi(x-i/2)\, x \frac{\partial}{\partial x}\, \phi(x+i/2) 
\ee
and
\be\label{eq:1st_derivative_filter}
d_i^{(00)} \equiv \int_\mathbb{R} \dd x \,\phi(x-i)\frac{\partial}{\partial x} \phi(x) = - d_{-i}^{(00)}\,.
\ee

The last equality is a consequence of the scaling functions and wavelets' compact support.
The results for $W^{(01)}_{m,n}$, $W^{(10)}_{m,n}$, $W^{(11)}_{m,n}$ follow analogously, provided that in the preceding
equations we change $\phi(x)\rightarrow\psi(x)$ and $0\rightarrow 1$ according to Eqs.\ \eqref{eq:W_matrixElements}.

The entries of the vector $d^{(00)}$ can be computed following the steps outlined in Ref.\ \onlinecite{Goedecker2009}  (Section 23), namely by exploiting the scaling equation featured by scaling functions and wavelets, Eqs.\ (\ref{eq:ref_rel}-\ref{eq:ref_rel2}).
Thanks to the localization of scaling functions, the number of non-trivial $d_j^{(00)}$'s amounts to just
a few: $d_j^{(00)} \neq 0 \Leftrightarrow |j|\leq 2m-2$, where $m$ is the order of the Daubechies wavelet family. The same holds for the other $d$ vectors, which, as a further consequence of  Eqs.\ (\ref{eq:ref_rel}-\ref{eq:ref_rel2}), can be computed as follows:
\begin{subequations}
\ba
d_i^{(01)} = 2\sum_{\mu, \nu} h_\mu\, g_\nu\,d^{(00)}_{2i+\mu-\nu}\;;\\
d_i^{(10)} = 2\sum_{\mu, \nu} g_\mu\, h_\nu\, d^{(00)}_{2i+\mu-\nu}\;;\\
d_i^{(11)} = 2\sum_{\mu, \nu} g_\mu\, g_\nu\, d^{(00)}_{2i+\mu-\nu}\;,
\ea
\end{subequations}
where $\{h_\mu\}_{\mu \in [1-m,m]}$, $\{g_\mu\}_{\mu \in [1-m,m]}$ are the low- and high-pass filters, respectively (cf.\ Sec.\ \ref{sec:Hamiltonian_WL}).

The entries of the vector $w^{(00)}$ can instead be computed as explained below, where $M\equiv (\mu+\nu)/2$,
$N\equiv (\mu-\nu)/2$, and the scaling relation Eq.~\eqref{eq:ref_rel} is used:
\begin{widetext}
\begin{subequations}
\ba
w^{(00)}_{J}
&=& 2 \sum_{\mu,\nu} h_\mu h_\nu \int \dd x\, \phi(2x-J-\mu)\, x \frac{\partial}{\partial x}\, \phi(2x+J-\nu)  \\
&=& 2 \sum_{M,N} h_{M+N} h_{M-N} \int \dd x\, \phi(2x-J-M-N)\, x \frac{\partial}{\partial x}\, \phi(2x+J-M+N)  \\
&=& \sum_{M,N} h_{M+N} h_{M-N} \int \dd x\, \phi(x-J-N)\, (x+M) \frac{\partial}{\partial x}\, \phi(x+J+N)  \\
&=&\sum_{M,N}  h_{M+N} h_{M-N}  \left[w^{(00)}_{2(J+N)}+M d^{(00)}_{2(J+N)} \right]\,. \label{eq:aJ_last_line}
\ea
\end{subequations}
\end{widetext}
Integrating by parts Eq.~\eqref{eq:a_J_def}, it is easy to prove that 
\begin{subequations}
\ba
w^{(00),(11)}_j &=& -w^{(00),(11)}_{-j} - \delta_{j,0}\;;\\
w^{(01)}_j &=& -w^{(10)}_{-j}\;,
\ea
\end{subequations}
and, as a corollary, that $w^{(00)}_0 = -1/2=w^{(11)}_0$. 
Moreover, $|j|> 2m-2 \Rightarrow w^{(00),(01),(10),(11)}_j = 0$.
Restoring the original indexes $\mu,\nu$ in \eqref{eq:aJ_last_line}, and rearranging terms in a convenient way, we are left with
\be
\label{eq:linear_system}
\sum_{k} A_{J,k} \, w^{(00)}_k = b_J\,,
\ee
where
\be\label{eq:linear_system-2}
A_{J,k} \equiv \delta_{J,k} - \sum_{\mu,\nu} h_\mu\, h_\nu\, \delta_{\mu-\nu+2J,k}
\ee
and
\be\label{eq:linear_system-3}
b_J \equiv  \sum_{\mu,\nu} h_\mu\, h_\nu \, \frac{\mu+\nu}{2} \,d^{(00)}_{\mu-\nu+2J}\,.
\ee
Namely, the entries of $w^{(00)}$ can be found as the solution of the linear system of equations \eqref{eq:linear_system}, which is neither under- nor over-determined. Formally equivalent linear systems, although with different numerical coefficients, lead to the solution for the vectors $w^{(01)}$, $w^{(10)}$, $w^{(11)}$. For instance, $w^{(01)}$ is the solution of 
\be
\sum_{k} A'_{J,k} \, w^{(01)}_k = b'_J\,,
\ee
with
\be
A'_{J,k} \equiv \delta_{J,k} - \sum_{\mu,\nu} h_\mu\, g_\nu\, \delta_{\mu-\nu+2J,k}
\ee
and
\be
b'_J \equiv  \sum_{\mu,\nu} h_\mu\, g_\nu \, \frac{\mu+\nu}{2} \,d^{(01)}_{\mu-\nu+2J}\,.
\ee


\bibliography{Cerioni_Genovese_Duchemin_Deutsch_032013}

\end{document}